\documentclass[aps,prd,preprintnumbers,superscriptaddress,nofootinbib,showpacs,twocolumn]{revtex4}%

\usepackage{amsmath,amssymb,bm,float,Macro,mathrsfs}
\usepackage[dvips]{graphicx}
\usepackage{color}
\usepackage{hyperref}

\def\grad{\phi}
\def\curl{\varpi}

\def\hC{\widehat{C}}
\def\hCEE{\widehat{C}^{\rm EE}}

\def\CEE{C^{\rm EE}}

\def\bn{\bm{\nabla}}

\let\normalcolor\relax

\begin{document}


\title{Future detectability of gravitational-wave induced lensing from 
high-sensitivity CMB experiments}

\author{Toshiya Namikawa}
\affiliation{Department of Physics, Stanford University, Stanford, CA 94305, USA}
\affiliation{Kavli Institute for Particle Astrophysics and Cosmology, SLAC National Accelerator
Laboratory, Menlo Park, CA 94025, USA}
\author{Daisuke Yamauchi}
\affiliation{Research Center for the Early Universe, School of Science, The University of Tokyo, Bunkyo-ku, Tokyo 113-0033, Japan}
\author{Atsushi Taruya}
\affiliation{Yukawa Institute for Theoretical Physics, Kyoto University, Kyoto 606-8502, Japan}
\affiliation{Kavli Institute for the Physics and Mathematics of the Universe, Todai Institutes for Advanced Study, 
the University of Tokyo, Kashiwa, Chiba 277-8583, Japan (Kavli IPMU, WPI)}

\date{\today}

\preprint{RESCEU-46/14,\,YITP-14-92}

\begin{abstract}
We discuss the future detectability of gravitational-wave induced lensing from high-sensitivity 
cosmic microwave background (CMB) experiments. Gravitational waves can induce a rotational component 
of the weak-lensing deflection angle, usually referred to as the curl mode, which would be imprinted 
on the CMB maps. Using the technique of reconstructing lensing signals involved in CMB maps, 
this curl mode can be measured in an unbiased manner, offering an independent confirmation of 
the gravitational waves complementary to B-mode polarization experiments. 
Based on the Fisher matrix analysis, we first show that with the noise levels necessary to confirm the 
consistency relation for the primordial gravitational waves, the future CMB experiments 
will be able to detect the gravitational-wave induced lensing signals. 
For a tensor-to-scalar ratio of $r\lsim 0.1$, even if the consistency relation is difficult to confirm 
with a high significance, the gravitational-wave induced lensing will be detected at more than 
$3\sigma$ significance level. 
Further, we point out that high-sensitivity experiments will be also powerful to constrain 
the gravitational waves generated after the recombination epoch. 
Compared to the B-mode polarization, the curl mode is particularly sensitive to gravitational waves 
generated at low redshifts ($z\lsim 10$) with a low frequency  ($k\lsim 10^{-3}\,$Mpc$^{-1}$), 
and it could give a much tighter constraint on their energy density $\Omega\rom{GW}$ 
by more than $3$ orders of magnitude. 
\end{abstract} 

\maketitle

\section{Introduction} \label{sec.1}

Large-scale B-mode polarizations of the cosmic microwave background (CMB) have been considered 
to be the powerful probe of the primordial gravitational waves as a smoking gun of the cosmic 
inflation. Recently, BICEP2 has reported a detection of B-mode signal consistent with primordial 
gravitational waves of the tensor-to-scalar ratio $r\sim 0.2$ \cite{Ade:2014xna}, though 
contaminations by the dust polarization seem significant \cite{Adam:2014bub}. Regardless of the 
origin of the signal, it is natural to explore further tests of the inflationary paradigm. 
Most inflation models predict a power-law form of the primordial tensor power spectrum. 
In particular, single-field slow-roll inflations provide a generic prediction, $n_{\rm t}=-r/8$\, 
known as the consistency relation, where $n_{\rm t}$ is the power-law index of the tensor spectrum. 
Indeed, confirmation of the consistency relation leads to a strong evidence for the accelerated 
expansion in the early Universe, and, thus, the next target after discovery of the primordial 
B-mode signal. 

Aiming at precisely measuring the tensor B-mode, next-generation CMB experiments are planning to 
achieve a high-sensitivity polarization measurement down to arc minute scales 
\cite{Abazajian:2013vfg,Abazajian:2013oma}. An important step in such future experiments is 
to reduce the lensing-induced B-mode with the so-called {\it delensing} technique. In CMB experiments, 
the gravitational potential of the large-scale structure can be reconstructed from the observed CMB 
maps, and the small-scale E/B-modes have been used to estimate the lensing signals involved in CMB 
maps by several CMB experiments such as SPTpol \cite{Hanson:2013hsb} and POLARBEAR \cite{Ade:2013gez}. 
The reconstructed lensing signals are then used to estimate the lensing B-mode to be subtracted from 
the observed B-mode map. Although this {\it delensing} technique requires both high-resolution and 
high-sensitivity experiments in practice, the Stage-IV class experiment 
\cite{Abazajian:2013vfg,Abazajian:2013oma} would have enough sensitivity to remove $\sim 80$-$90$\% of 
the lensing B-mode, giving us a chance to confirm the consistency relation if $r\sim 0.2$ 
\cite{Boyle:2014kba,Simard:2014aqa}. An important point is that such experiments will not only 
greatly improve our understanding of the physics in the early Universe, but also give another 
benefit for the detection of tensor fluctuations through the lensing effect, which we will focus on. 

In general, the weak-lensing deflection angle is decomposed into the (parity-even) gradient and 
(parity-odd) curl modes, expressed as \cite{Hirata:2003ka,Cooray:2005hm,Namikawa:2011cs} 
\beq 
	\bm{d} = \bm{\nabla}\grad + (\star \bm{\nabla})\curl 
	\,. \label{deflection}
\eeq 
The quantities $\grad$ and $\curl$, respectively, denote the gradient and curl modes of the deflection 
angle. The symbol $\bn$ is the covariant derivative on the unit sphere, and the operator $\star$ 
rotates a two-dimensional vector counterclockwise by $90^{\circ}$. 
Similar to the cases of E-/B-mode CMB polarizations, the scalar metric perturbations induced by the 
matter density fluctuations produce the gradient mode, but they do not generate the curl mode 
at the linear order. On the other hand, the curl mode is generated by the vector and/or tensor metric 
perturbations, and is, thus, considered as an alternative probe of the primordial gravitational waves 
\cite{Dodelson:2003bv,Cooray:2005hm,Li:2006si,Namikawa:2011cs}. 

In previous works, the detectability of the curl mode has been discussed based on the lensing 
reconstruction with the quadratic estimator \cite{Cooray:2005hm,Namikawa:2011cs}, and the detection of 
the gravitational-wave induced lensing is found to be difficult for $r\approx 0.2$ even with the 
cosmic-variance limited observation. Note, however, that their results do not imply the fundamental 
limit of the detectability of the primordial gravitational waves, because the reconstruction noise 
is further reduced if the estimator based on the maximum likelihood method is used 
\cite{Hirata:2003ka}. The primary purpose of this paper is to reconsider the detectability of 
the curl mode from the primordial gravitational waves based on the maximum likelihood approach. 
We will show that, with a high-sensitivity experiment enough to confirm the consistency relation, 
the detection of the gravitational-wave induced lensing is possible with a high signal-to-noise ratio. 

There is another benefit to search for the gravitational-wave induced lensing. Indeed, 
the gravitational waves of cosmological origin can be also generated at late time of the Universe. 
Possible mechanisms to generate gravitational waves at the late-time epoch include 
second-order primordial 
density perturbations \cite{Baumann:2007zm}, cosmic strings \cite{Yamauchi:2013fra}, 
anisotropic stress of a scalar field in modified gravity theories (see, e.g., \cite{Saltas:2014dha}), 
and self-ordering scalar fields \cite{Fenu:2009JCAP}(more generically, any cosmic defect network 
\cite{Figueroa:2013PRL}). 
In order to 
constrain these gravitational waves, we need a sensitive probe to the late-time evolution of 
the gravitational waves. CMB lensing measurements would provide a way to detect those gravitational 
waves, since the lensing effect on CMB is efficient at late time of the Universe. The postrecombination 
gravitational waves, if they exist, give a small but nonvanishing contribution to the 
lensing effect on the observed CMB anisotropies \cite{2012:Rotti}. While the possibility to constrain 
the postrecombination gravitational waves using the lensed CMB angular power spectra has been 
considered by Ref.~\cite{2012:Rotti}, the direct reconstruction of the curl mode will further improve 
the constraint, since the reconstruction utilizes information on the statistical anisotropies solely 
caused by the lensing. In this paper, we will show that even if the sensitivity of an experiment is 
marginal to detect the curl mode of the primordial gravitational waves, the constraint on the post-
recombination gravitational waves can become tighter than that from the B-mode polarization by more 
than $3$ orders of magnitude.

This paper is organized as follows. In Sec~\ref{sec.2}, we begin by briefly summarizing our basic 
method to estimate the efficiency of the lensing reconstruction and delensing based on the maximum 
likelihood method. Then, in Sec.~\ref{sec.3}, we discuss prospects for a precision measurement of the 
primordial gravitational waves via high-sensitivity CMB experiments, showing a capability of 
detecting gravitational-wave induced lensing. In Sec.~\ref{sec.4}, as an implication to the search 
for the gravitational-wave induced lensing, we consider gravitational waves produced after the 
recombination epoch, and discuss the possibility to constrain the energy density of the 
postrecombination gravitational waves via the CMB lensing analysis. 
Finally, Sec~\ref{sec.5} is devoted to summary and discussion.

Throughout this paper, we assume the flat-$\Lambda$CDM model with three massless neutrinos as 
the fiducial cosmological model. The cosmological parameters used in this paper are consistent with 
the best-fit values of the Planck 2013 results \cite{Ade:2013zuv}:  
$\Omega\rom{b}h^2=0.0220$, $\Omega\rom{m}h^2=0.141$, $\Omega_{\Lambda}=0.696$, 
$n\rom{s}=0.968$, $A\rom{s}=2.22\times10^{-9}$, and $\tau=0.0949$. 
The pivot scale of the primordial scalar/tensor power spectrum is $k_0=0.05$ Mpc$^{-1}$. 
For tensor fluctuations, we take the fiducial values of the tensor tilt to follow the consistency 
relation $n_{\rm t}=-r/8$, while the fiducial tensor-to-scalar ratio $r$ is varied.

\section{Analysis of CMB lensing} \label{sec.2}

In this section, we summarize our basic method to estimate the efficiency of the lensing 
reconstruction and delensing from CMB experiments, used to quantify the sensitivity to 
gravitational-wave induced B-mode and lensing in the subsequent analysis.

\subsection{Angular power spectrum} 

In what follows, we use {\tt CAMB} code \cite{Lewis:1999bs} to compute angular power spectra 
of the E-/B-mode polarizations and gradient/curl modes, following our previous work 
\cite{Yamauchi:2013fra}. Denoting the dimensionless power spectrum for tensor perturbations by 
$\Delta_h^2(k,\eta)$\,, angular power spectra induced by the gravitational waves are expressed as:
\al{
	& C^{\rm XX}_\ell
		= 4\pi\INT{}{\ln k}{_}{0}{\infty} \biggl[\INT{}{\chi}{_}{0}{\infty} \Delta_h 
			(k,\eta_0 -\chi )S_{{\rm X},\ell}(k,\chi )\biggr]^2
	\,,
}
where the index ${\rm X}$ runs over the E-/B-mode polarizations and the lensing gradient-/curl-modes 
($\rm X=E,B,\grad$ or $\curl$). The function $S_{{\rm X},\ell}$ is the weight function for ${\rm X}$\, 
(see Ref.~\cite{Yamauchi:2013fra} for explicit expressions). In particular, the weight function for 
the curl mode is given by 
\al{
	& S_{\varpi,\ell} = \frac{1}{2}\frac{(\ell -1)!}{(\ell +1)!}
		\sqrt{\frac{(\ell +2)!}{(\ell -2)!}}\frac{j_\ell (k\chi )}{k\chi^2}
	\,. \label{eq:weight_curl}
}
For the evolution of the tensor perturbations, we introduce the tensor transfer function $T(k,\eta )$, 
which basically describes the subhorizon evolution of gravitational waves. In terms of this, 
we may write the time evolution of the power spectrum as 
$\Delta_h^2 (k,\eta)=\Delta_{h,{\rm prim}}^2 (k)[T(k,\eta)]^2$. 
Here $\Delta_{h,{\rm prim}}^2 (k)$ denotes the dimensionless power spectrum of the primordial 
tensor perturbations produced during inflation, which is characterized by the tensor-to-scalar ratio 
$r$ and tensor tilt $n_{\rm t}$ through $\Delta_{h,{\rm prim}}^2 (k)=rA_{\rm s}(k/k_0)^{n_{\rm t}}$ 
with $A_{\rm s}$ being the power spectrum amplitude of the scalar perturbations.

In our analysis, we assume that an instrumental noise of observed polarization maps 
is homogeneous and isotropic, and is deconvolved with a Gaussian beam. 
The E-/B-mode polarization power spectra are then given by 
\al{
	\hC^{\rm XX}_{\ell} = C^{\rm XX}_{\ell} + \left(\frac{\sigma\rom{P}}{T\rom{CMB}}\right)^2
		\exp\left[\frac{\ell(\ell+1)\theta^2}{8\ln 2}\right]
	\,, \label{eq:hat C^XX}
}
where $\rm XX=EE$ or $\rm BB$, the quantities, $T\rom{CMB}$, $\theta$, and $\sigma\rom{P}$ 
are, respectively, the mean temperature of CMB (i.e., $T\rom{CMB}=2.7$K), the beam size, and 
the noise level of an experiment. 
Note that the power spectrum $C^{\rm XX}_{\ell}$ is the lensed 
E-/B-mode polarization power spectra. On top of the primordial component, it includes the lensing 
contributions.

\subsection{Lensing reconstruction and delensing} 

Provided the observed polarization maps, Ref.~\cite{Hirata:2003ka} has proposed an iterative method to 
reconstruct the gradient and curl modes of the deflection angle based on the maximum likelihood 
method. While this method enables us to substantially reduce the reconstruction noise, 
the estimation of the noise generally requires extensive numerical simulations.  
In Ref.~\cite{Smith:2010gu}, instead of performing full numerical simulations, a fast and simple 
algorithm to estimate the reconstruction noise is exploited for both the gradient mode and the 
delensed B-mode polarization, and it has been used for several forecast studies 
\cite{Abazajian:2013vfg,Abazajian:2013oma,Wu:2014hta,Boyle:2014kba,Simard:2014aqa}. 
In this paper, extending the forecast method by Ref.~\cite{Smith:2010gu} to 
the case including the curl mode, we will estimate the expected reconstruction noise for the gradient 
and curl modes as follows. 
Note that the primordial tensor contributions to the E-/B-mode polarizations 
are basically small enough and are irrelevant in estimating the reconstruction noise, 
since these are dominated only at larger scales ($\ell\lsim 500$) if $r\lsim 0.2$. 
Hence, in the lensing analysis, we basically follow the previous studies and 
ignore the tensor contributions to the polarizations.  

In the iterative method, as a first step, the lensing reconstruction is performed with the usual 
quadratic estimator \cite{Okamoto:2003zw}. The reconstruction noise of the gradient mode is then 
given by \cite{Smith:2010gu} 
\al{
	N_L^{\grad} &= \left\{\frac{1}{2L+1}\sum_{\ell\ell'}
		\frac{[\mC{S}^{(-)}_{\ell'\ell L}\CEE_{\ell}]^2}
		{\hC^{\rm EE}_{\ell}\hC^{\rm BB}_{\ell'}} \right\}^{-1}
	\,. \label{Eq:Nlgg}
}
Here we define 
\al{ 
	\mC{S}^{(\pm)}_{\ell\ell'L} 
		&= \frac{1\pm (-1)^{\ell+\ell'+L}}{2} 
			\sqrt{\frac{(2\ell+1)(2\ell'+1)(2L+1)}{16\pi}}
	\notag \\ 
		&\qquad \times 
			[-\ell(\ell+1)+\ell'(\ell'+1)+L(L+1)]
	\notag \\ 
		&\qquad \times 
		\Wjm{\ell}{\ell'}{L}{2}{-2}{0} 
	\,. 
} 
With the reconstructed gradient mode, the lensing contributions in the B-mode polarization 
are estimated and are then subtracted from the observed B-mode polarization. 
Because of the imperfect subtraction, there remain the residuals of the lensing contaminations 
in the B-mode polarization. The residual lensing contribution to the 
angular power spectrum is estimated to be \cite{Smith:2010gu} 
\al{
	C^{\rm BB,res}_{\ell} &= \sum_{\ell' L}\frac{\CEE_{\ell'}C_L^{\grad\grad}}{2\ell+1} 
		\left[1-(\mC{S}^{(-)}_{\ell\ell'L})^2 \frac{\CEE_{\ell'}}{\hCEE_{\ell'}}
			\frac{C_L^{\grad\grad}}{C_L^{\grad\grad}+N_L^{\grad}}\right]
	\,. \label{Eq:rClBB}
}
Note that, in the above equation, we ignore the B-mode polarization arising from 
the gravitational-wave induced lensing, and we assume that the lensing B-mode polarization primarily 
comes from the gradient mode. 

The next step is to calculate the reconstruction noise of Eq.~(\ref{Eq:Nlgg}) again, 
taking the residual lensing contributions into account. 
This can be done by replacing the lensing contribution in 
$\hC^{\rm BB}_{\ell}$ with $C_{\ell}^{\rm BB,res}$. Similarly, the lensing contribution in 
$\hC^{\rm EE}_{\ell}$ may be replaced with the residual lensing contribution; however, the relative 
impact of the lensing contribution is rather small and is safely ignored in the E-mode 
polarization. Then, we estimate the new residual contribution to the B-mode polarization. 
We repeat these calculations until the power spectrum $C_\ell^{\rm BB,res}$ is converged. 
The converged result of $C_\ell^{\rm BB,res}$ would be regarded as the final outcome of 
the residual lensing contamination after delensing \cite{Smith:2010gu}. 
With this result, we obtain the reconstruction noise of the curl mode as follows:
\al{
	N_L^{\curl} &= \left\{\frac{1}{2L+1}\sum_{\ell\ell'}
		\frac{[\mC{S}^{(+)}_{\ell'\ell L}\CEE_{\ell}]^2}
			{\hC^{\rm EE}_{\ell}\hC^{\rm BB}_{\ell'}} \right\}^{-1}
	\,, \label{Eq:Nlcc}
}
where the lensing contributions in $\hC^{\rm BB}_{\ell}$ are replaced with the converged result of 
$C^{\rm BB,res}_{\ell}$. In our subsequent analysis, the multipoles between 
$2\leq\ell,\ell',L\leq 4000$ are used for the reconstruction and delensing in 
Eqs.~\eqref{Eq:Nlgg}, \eqref{Eq:rClBB}, and \eqref{Eq:Nlcc}. 

The iteration method given above implicitly assumes that the correlation between the gradient and 
curl modes is negligible; i.e., the gradient and curl modes are estimated separately. 
Although such correlation for the maximum likelihood reconstruction has not been explored in detail, 
their impact on our results would be not so significant because the assumption is, indeed, true for 
the quadratic estimator \cite{Namikawa:2011cs} which is used as a first step of the iteration.

\section{Measuring primordial gravitational waves via high-sensitivity CMB experiment} 
\label{sec.3}

In this section, we first estimate the required noise level of a polarization measurement 
in order to test 
the consistency relation for primordial tensor fluctuations. Based on this estimate, we discuss 
the feasibility to detect the gravitational-wave induced lensing from a high-sensitivity CMB 
experiment. 

\subsection{Testing consistency relation with B-mode polarization} 
\label{sec.3.2}

\begin{figure*}
\bc
\includegraphics[height=70mm,clip]{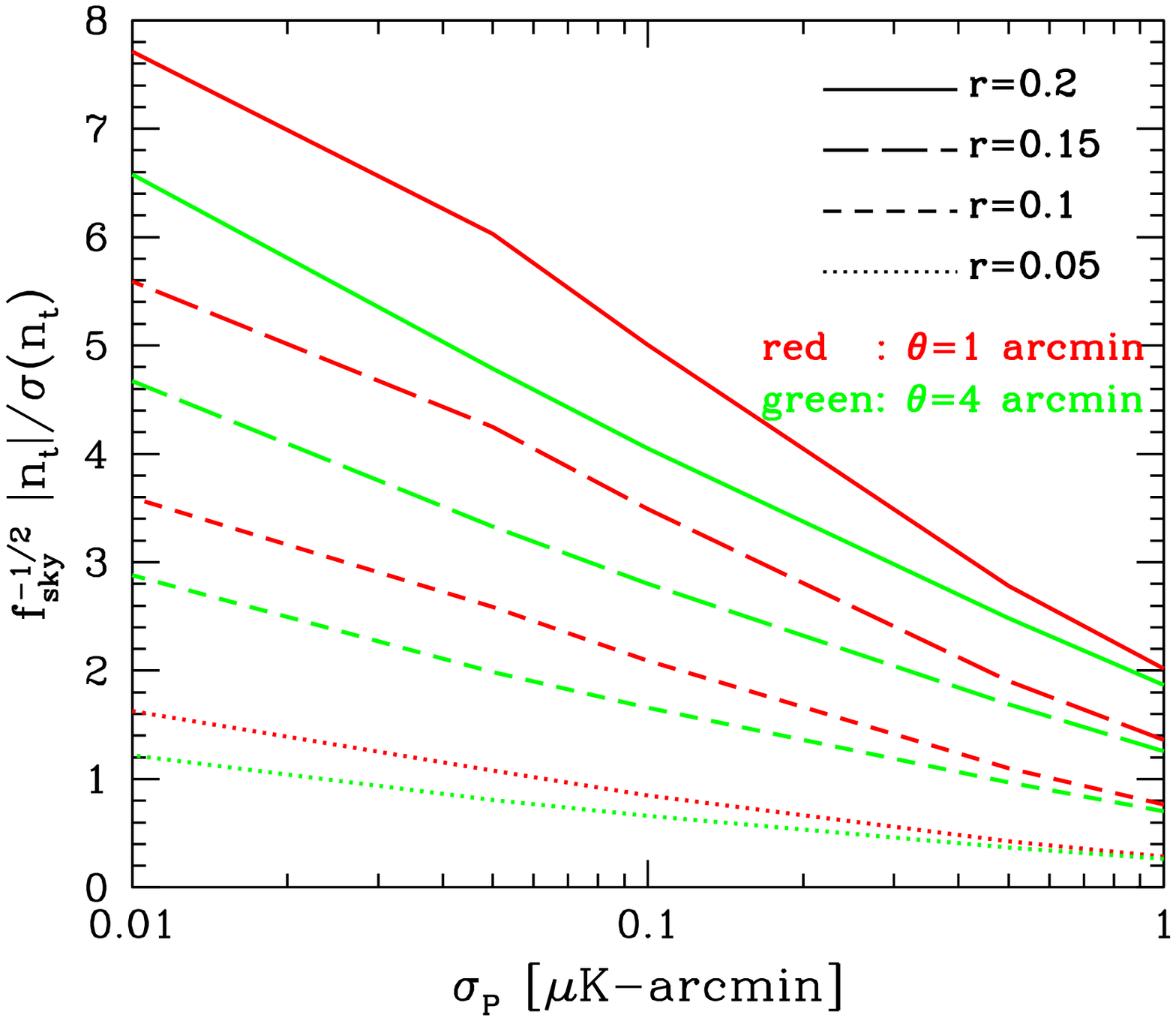}
\includegraphics[height=70mm,clip]{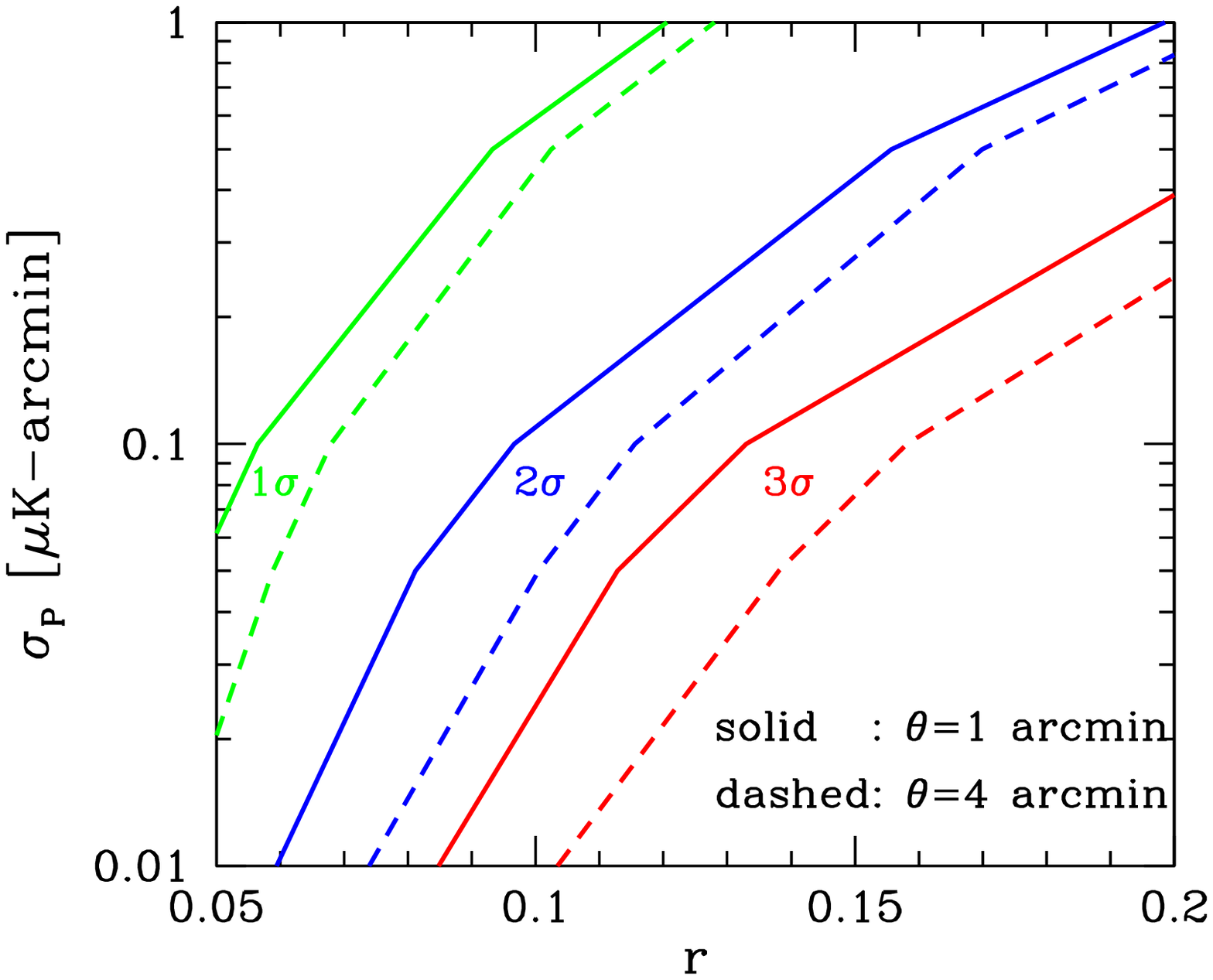}
\caption{
{\it Left}: 
Statistical significance to confirm the consistency relation, 
$f_{\rm sky}^{-1/2}\,|n_{\rm t}|/\sigma(n\rom{t})$, with the fiducial value of the spectral tilt 
given by $n_{\rm t}=-r/8$. The results for tensor-to-scalar ratios $r=0.2$ (solid), 
$0.15$ (long dashed), $0.1$ (short dashed), and $0.05$ (dotted) are plotted as a function of 
the polarization sensitivity $\sigma\rom{P}$ in units of $\mu$K-arcmin. The red and green lines, 
respectively, represent angular resolutions of $\theta=1$ and $4$ arcmin. 
{\it Right}: 
Required polarization sensitivity to test the consistency relation as 
a function of the tenor-to-scalar ratio $r$ for a full-sky CMB experiment ($f_{\rm sky}=1$). 
For the angular resolution $\theta=1$ (solid) and $4$ arcmin (dashed), the required sensitivity 
$\sigma_{\rm P}$ is shown at the statistical significance of $1\sigma$ (green), $2\sigma$ (blue), 
and $3\sigma$ (red) levels.  
}
\label{expected}
\ec
\end{figure*}

\begin{figure*}
\bc
\includegraphics[height=70mm,clip]{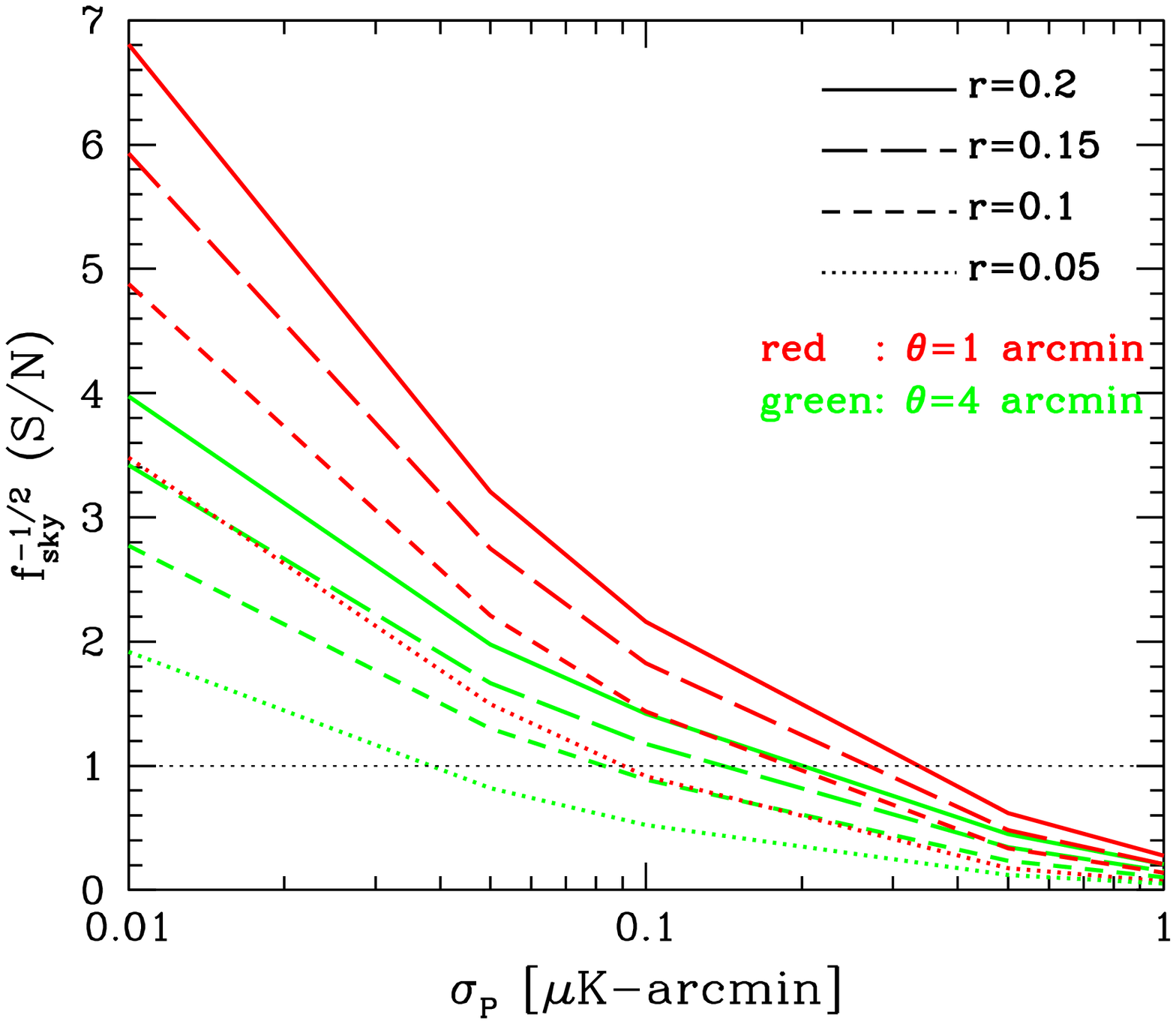}
\includegraphics[height=70mm,clip]{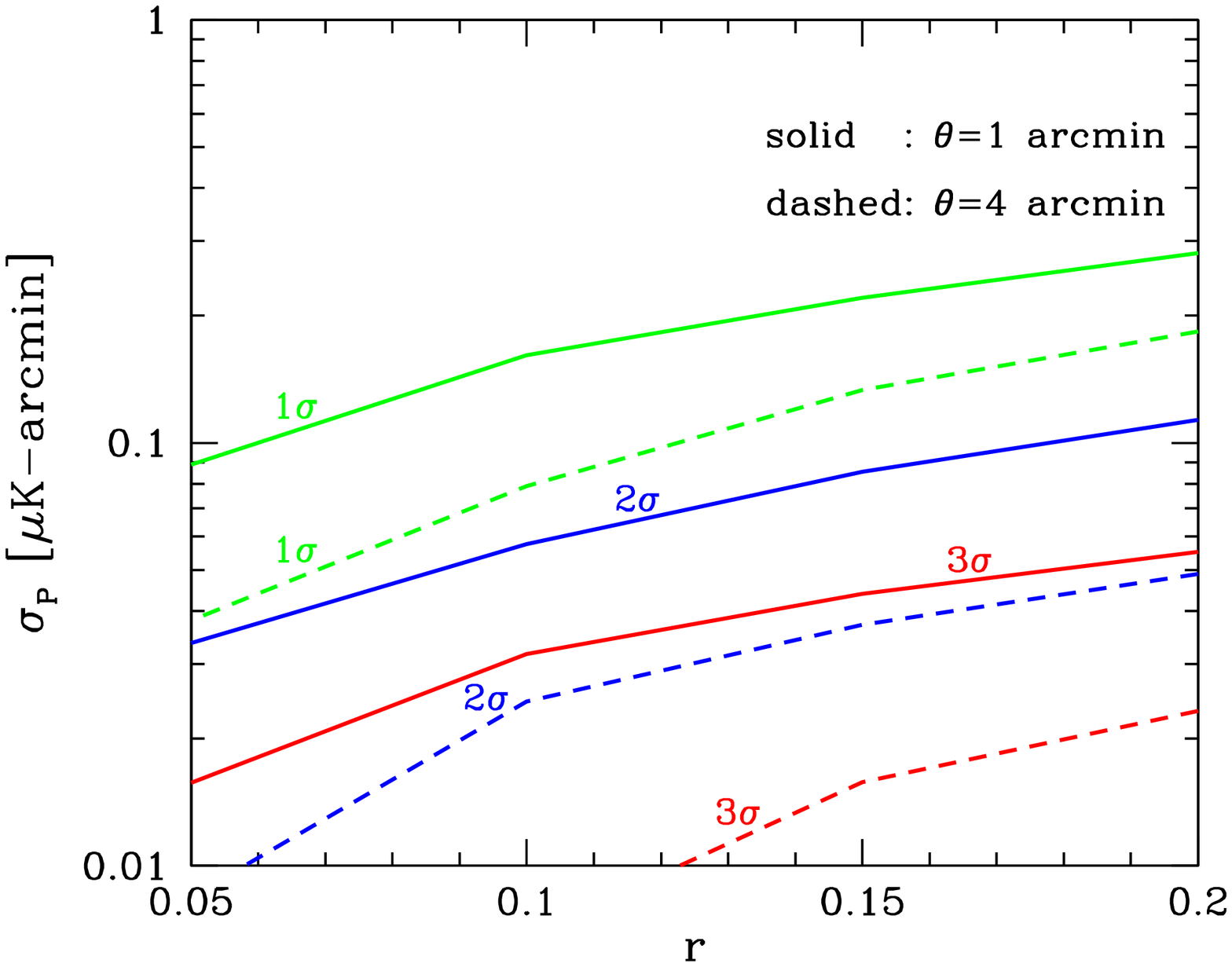}
\caption{
{\it Left}: 
Signal-to-noise ratio of the lensing curl-mode induced by the primordial gravitational waves 
$f_{\rm sky}^{-1/2}\,\mbox{(S/N)}_{\curl\curl}$ plotted against the sensitivity $\sigma\rom{P}$. 
The meaning of the line types is the same as in the left panel of Fig.~\ref{expected}. 
As a reference, the $1\sigma$ significance is shown in as the black solid line.  
{\it Right}: 
Required sensitivity to detect the lensing curl-mode at $1\sigma$ (green), $2\sigma$ (blue), 
and $3\sigma$ (red) levels. For a full-sky CMB experiment ($f_{\rm sky}=1$), the results are 
plotted as a function of the tenor-to-scalar ratio $r$. The sold and dashed lines represent the results 
with angular resolutions of $\theta=1$ and $4$ arcmin, respectively. 
}
\label{snr_curl}
\ec
\end{figure*}

In order to derive the required noise level of a polarization measurement needed to test
the consistency relation $n\rom{t}=-r/8$, we proceed to the Fisher matrix analysis. 
Here we consider the B-mode polarization alone, since this is the best sensitive probe to test the 
consistency relation among various CMB observables. The Fisher matrix is given by 
\al{
	F_{ij} \equiv \sum_{\ell=2}^{\ell_{\rm max}}
		\frac{2\ell+1}{2}f\rom{sky}
		\PD{\ln \hC^{\rm BB}_{\ell}}{p_i}\PD{\ln \hC^{\rm BB}_{\ell}}{p_j}
	\,, 
        \label{eq:Fisher_ij}
}
with $\hC^{\rm BB}_\ell$ being the observed power spectrum given by Eq.~\eqref{eq:hat C^XX}, whose 
lensing contributions are replaced with Eq.~\eqref{Eq:rClBB}. For simplicity, we ignore the galactic 
foreground contamination in computing the power spectrum. The quantities $p_i$ are free parameters to 
be determined by the observations, and we here consider the tensor-to-scalar ratio and tensor spectral 
index as free parameters, i.e., $p_i=(r,n\rom{t})$. Given the Fisher matrix above, the marginalized 
expected $1\sigma$ error on $p_i$ is estimated to be $\sigma(p_i)=\sqrt{\{F^{-1}\}_{ii}}$. 
Below, setting the maximum multipole to $\ell_{\rm max}=1000$, we evaluate the Fisher matrix.  

It is commonly known that confirmation of the consistency relation generally requires a polarization 
measurement with a significantly low-noise level. We, thus, consider the Stage-IV class or even 
higher-sensitivity experiments in which the polarization sensitivity 
$\sigma_{\rm P}\lesssim 1\,\mu$K-arcmin 
and angular resolution $\theta\lesssim 1$ arcmin can be achieved. For comparison, we also show the 
case with $\theta=4$ arcmin which is close to the angular resolution of the POLARBEAR and POLAR Array. 
With these experimental setups, we apply the delensing technique, and estimate 
the expected error on the tensor spectral tilt from the delensed B-mode polarization 
based on Eq.~\eqref{eq:Fisher_ij}. 
Note that, throughout the analysis, we assume the consistency relation for setting the fiducial 
value of $n\rom{t}$. 

The left panel of Fig.~\ref{expected} shows the statistical significance to confirm the consistency 
relation, defined by $|n_{\rm t}|/(\sqrt{f_{\rm sky}}\sigma (n_{\rm t}))$ with the fiducial value of 
the tensor spectral tilt, $n_{\rm t}=-r/8$. For the tensor-to-scalar ratios of $r=0.2$ (solid), 
$0.15$ (long dashed), $0.1$ (short dashed), and $0.05$ (dotted), statistical significances are 
estimated and are plotted as a function of the sensitivity $\sigma_{\rm P}$. Note that thanks to 
the low noise level of a polarization measurement, the tensor-to-scalar ratio is tightly constrained 
for the range of our interest in $r$, and, thus, the detection of primordial tensor perturbations is 
highly significant, i.e., $r/(\sqrt{f_{\rm sky}}\sigma(r))\gg 1$. 

As it is anticipated, the confirmation of the consistency relation becomes harder as decreasing 
the fiducial value of $r$. This is true even with a full-sky ($f_{\rm sky}=1$) and 
high-angular resolution ($\theta=1$ arcmin) experiment. For instance, for $r<0.05$, the sensitivity 
of $\sigma_{\rm P}\ll 0.1\,\mu$K-arcmin is required to confirm the consistency relation. 
This is more clearly seen when we plot the required sensitivity $\sigma_{\rm P}$ as a function of 
the tensor-to-scalar ratio. 
The right panel of Fig.~\ref{expected} plots the required sensitivity based on 
the result in the left panel. Here, assuming full-sky observations ($f_{\rm sky}=1$), 
the results with statistical significance at the $1\sigma$ (green), $2\sigma$ (blue), and 
$3\sigma$ (red) levels are particularly shown. For a small tensor-to-scalar ratio $r\lesssim0.1$, 
a solid confirmation of the consistency relation at the $\gtrsim3\sigma$ level is rather challenging 
even for a high-sensitivity experiment of $\sigma_{\rm P}\gtrsim0.01\,\mu$K-arcmin.

\subsection{Detecting gravitational-wave induced lensing} \label{sec.3.3} 

Having confirmed that testing the consistency relation generally requires a high-sensitivity B-mode 
measurement, we next consider the feasibility to detect the gravitational-wave induced lensing. 
We estimate the expected signal-to-noise ratio for the reconstructed curl-mode. 
The signal-to-noise ratio of the curl mode is defined as 
\al{
	\left(\frac{\rm S}{\rm N}\right)_{\curl\curl} 
		\equiv \left[\sum_{\ell=2}^{\ell_{\rm max}} 
		\frac{2\ell + 1}{2}f\rom{sky} 
		\left(\frac{C_{\ell}^{\curl\curl}}{\hC_{\ell}^{\curl\curl}}\right)^2 
		\right]^{1/2} \,. \label{SN}
}
Note that the observed power spectrum of the curl mode $\hC^{\curl\curl}_{\ell}$ is expressed as 
\al{
	\hC^{\curl\curl}_{\ell} = C_{\ell}^{\curl\curl} + N_{\ell}^{\curl}
	\,,\label{eq:hat C^curl}
}
with the reconstruction noise $N_\ell^\curl$ given by Eq.~\eqref{Eq:Nlcc}.

The left panel of Fig.~\ref{snr_curl} shows the expected signal-to-noise ratio for the curl mode as a 
function of the sensitivity $\sigma_{\rm P}$. On the other hand, right panel of Fig.~\ref{expected} 
plots the required sensitivity to detect the curl mode as a function of $r$. 
Here we set $\ell_{\rm max}=1000$ 
\footnote{In practice, the signal-to-noise ratio is well-converged if we set $\ell_{\rm max}$ to 
several hundreds}. 
Similar to the test of the consistency relation, the detectability of the curl mode increases with 
the fiducial value of $r$. 
However, one noticeable point is that the required sensitivity for 
the detection of the gravitational-wave induced lensing is less severe than that for the test of 
the consistency relation. Indeed, the gravitational-wave induced lensing for $r\gsim 0.1$ will be 
detected with $3\sigma$ significance if the sensitivity of $\sigma_{\rm P}=0.03\,\mu$K-arcmin is 
achieved in future CMB experiments. In particular, even if the consistency relation is 
still difficult to probe, the curl mode would be detected for $r<0.1$. 
This implies that the high-sensitivity measurement of the curl mode can be a complementary probe 
of the primordial gravitational waves, offering an independent confirmation from future CMB 
experiments.

\section{Implication to postrecombination gravitational waves} \label{sec.4}

\begin{figure*}
\bc
\includegraphics[width=200mm,clip]{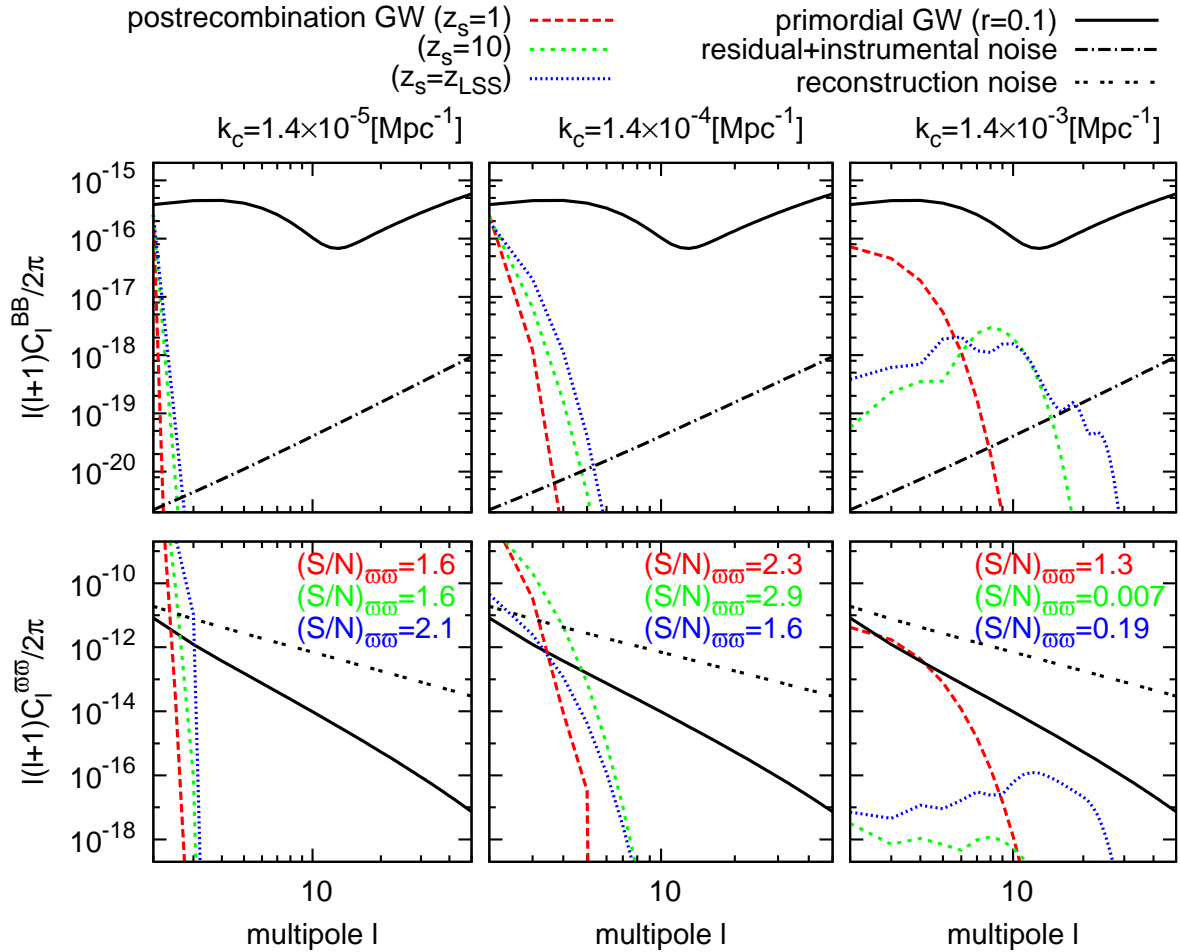}
\caption{
Angular power spectra of the B-mode polarization (top) and lensing curl-mode (bottom) induced by 
the postrecombination gravitational waves. For specific source redshifts at $z\rom{s}=1$ (red), 
$10$ (green) and $z_{\rm LSS}=1100$ (blue), the expected signals of the B-mode polarization and 
curl mode are shown. Left, middle, and right panels, respectively, show the results from the 
postrecombination gravitational waves produced at the frequencies $k\rom{c}=1.4\times10^{-5}$, 
$1.4\times10^{-4}$, and $1.4\times10^{-3}$\,\, Mpc$^{-1}$ (see the text for details). 
The amplitude of the gravitational waves given by $\Delta_{\rm GW}^2$ is normalized in each case 
such that the measured B-mode power spectrum has $(\mbox{S/N})_{\rm BB}=1$. For reference, 
we also plot in each panel the angular power spectra from the primordial gravitational waves,  
assuming the tensor-to-scalar ratio, $r=0.1$ (black solid). The noise contributions from the 
instrument and the lensing reconstruction or delensing are also depicted as dot-dashed lines 
in each panel.
}
\label{Cl_lowz}
\ec
\end{figure*}

As another benefit to search for the gravitational-wave induced lensing via a high-sensitivity CMB 
experiment, we here discuss the gravitational waves generated after the recombination. 
As we mentioned in Sec.~\ref{sec.1}, there are several mechanisms to produce 
the late-time gravitational waves: the second-order primordial density perturbations 
\cite{Baumann:2007zm} and anisotropic stress of a scalar field in modified gravity theories 
(see, e.g., \cite{Saltas:2014dha}). While the postrecombination gravitational waves lead additional 
features in both the angular power spectra of the B-mode polarization and curl mode, the curl mode 
is found to be particularly sensitive to the late-time gravitational waves on large scales. 

For illustrative purposes to show how the B-mode polarization and curl mode are sensitive to 
the postrecombination gravitational waves, we consider gravitational waves emitted instantaneously 
at a single source plane of $z=z\rom{s}$, with a narrow frequency range, 
$k\rom{c}(1-\epsilon )\leq k\leq k\rom{c}(1+\epsilon )$. In what follows, we set $\epsilon =0.28$. 
Adopting the same transfer function as used in the case of the primordial gravitational waves, 
we compute additional contributions to the angular power spectrum from the postrecombination 
gravitational waves denoted by $\Delta C^{\rm XX}_\ell$, as 
\al{
	\Delta C^{\rm XX}_\ell
		=& 4\pi\int^{k\rom{c}(1+\epsilon )}_{k\rom{c}(1-\epsilon )}{\rm d}\ln k\,\Delta_{\rm GW}^2
	\notag\\
	&\times
		\biggl[
			\int_0^{\chi (z\rom{s})}{\rm d}\chi\, T(k,\eta_0 -\chi )S_{{\rm X},\ell}(k,\chi )
		\biggr]^2
	\,.
}
Here, $T$ is the transfer function and $S_{{\rm X},\ell}$ represents the weight function 
[e.g., Eq.~(\ref{eq:weight_curl})]. In the above, the quantity $\Delta_{\rm GW}$ represents the amount 
of the gravitational waves produced at $z_s$, and it is treated as a free parameter. 

Fig.~\ref{Cl_lowz} shows the angular power spectra of the B-mode polarization (top) and curl mode  
(bottom) from the postrecombination gravitational waves with three different frequencies: 
$k_{\rm c}=1.4\times10^{-5}$ (left), $1.4\times10^{-4}$ (middle), and 
$1.4\times10^{-3}$ (right)\,\, Mpc$^{-1}$. Here, the amplitude of the gravitational waves 
$\Delta_{\rm GW}$ is normalized in each case in such a way that the expected signal-to-noise ratio 
of the B-mode polarization $(\mbox{S/N})_{\rm BB}$ becomes unity 
\footnote{
The signal-to-noise ratio $(\mbox{S/N})_{\rm BB}$ is obtained from Eq.~\eqref{SN} but replacing 
$C_{\ell}^{\curl\curl}$ and $\hC_{\ell}^{\curl\curl}$ with $C_{\ell}^{\rm BB}$ and 
$\hC_{\ell}^{\rm BB}$, respectively. 
}. 

Overall, the long-wavelength gravitational waves produce a very sharp feature in 
the angular power spectra. As indicated in Fig.~\ref{Cl_lowz}, 
although general trends of the power spectra are rather similar in 
both the B-mode polarization and lensing curl-mode, resultant signal-to-noise ratios of the curl mode 
$(\mbox{S/N})_{\curl\curl}$ are higher than those of the B-mode polarization especially for 
smaller $k\rom{c}$ and lower $z\rom{c}$. This suggests that the curl 
mode gives a better constraint on the abundance of the postrecombination gravitational waves.  

In order to quantify the potential power of the high-sensitivity CMB experiment to constrain 
the postrecombination gravitational waves, we estimate the expected $1\sigma$ upper bounds on 
$\Delta\rom{GW}^2$ based on the following log-likelihood: 
\al{
	- 2\ln\mC{L} 
		&= \sum_{\ell=2}^{\ell_{\rm max}}\, (2\ell+1)
	\notag \\ 
	&\qquad \times \bigg[\frac{\widehat{\mC{C}}_{\ell}^{\rm XX}}
          {\widetilde{\mC{C}}_{\ell}^{\rm XX}}
	+ \ln \widetilde{\mC{C}}_{\ell}^{\rm XX} 
		- \frac{2\ell-1}{2\ell+1}\ln \widehat{\mC{C}}_{\ell}^{\rm XX} \bigg]
	\,. 
}
Here the {\it observed} power spectra $\widehat{\mC{C}}_{\ell}^{\rm XX}$ are obtained by 
performing the lensing analysis, and we assume that these have no contribution 
from the postrecombination sources. 
On the other hand, $\widetilde{\mC{C}}_{\ell}^{\rm XX}$ is a {\it theoretically-estimated} 
power spectrum, which includes the contribution from the postrecombination gravitational waves 
$\Delta C_{\ell}^{\rm XX}$ on top of the observed power spectrum $\widehat{\mC{C}}_{\ell}^{\rm XX}$. 
In computing the likelihood above, we set the maximum multipole to $\ell_{\rm max}=1000$. 
The $1\sigma$ constraint on the amplitude $\Delta_{\rm GW}$ is then obtained for the range of 
frequency centered at $k_c$, $k\rom{c}(1-\epsilon )\leq k\leq k\rom{c}(1+\epsilon )$. 
For convenience, we translate this constraint into the upper bounds on the spectral energy density at 
the present time, frequently used in the literature:
\al{
	\Delta\Omega\rom{GW}(k) \equiv \frac{1}{\rho\rom{c}}\D{\Delta\rho\rom{GW}}{\ln k}
			\biggl|_{\eta =\eta_0}
		= \frac{\Delta\rom{GW}^2}{12H_0^2}
		\left(\PD{T(k,\eta )}{\eta}\right)^2\bigg|_{\eta =\eta_0} 
	\,,\label{eq:Omega_GW}
}
where $\rho\rom{c}$ is the critical density of the Universe, and $H_0$ is the Hubble parameter today.

\begin{figure*}
\bc
\includegraphics[width=170mm,clip]{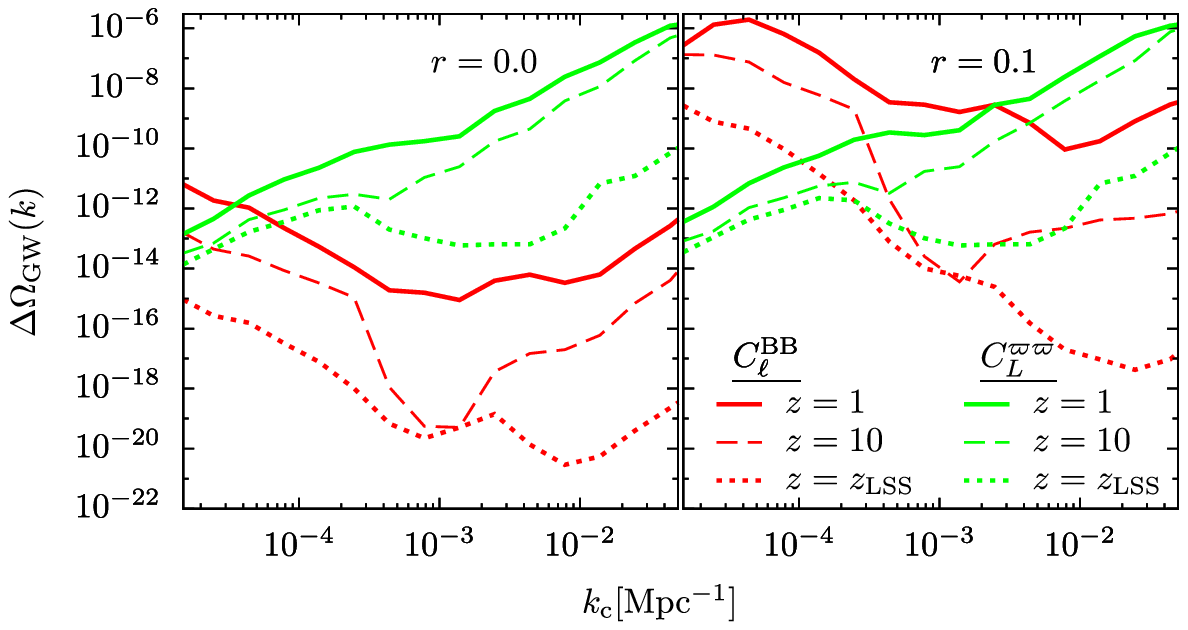}
\caption{
Upper bounds on the spectral energy density of the postrecombination gravitational waves 
$\Delta\Omega\rom{GW}$ as a function of the frequency. The left panel shows the results in the absence 
of the primordial gravitational waves $(r=0)$, while right panel represents the cases with 
nonzero primordial gravitational waves of the tensor-to-scalar ratio, $r=0.1$. In each panel, 
the constraints from the B-mode polarization and lensing curl-mode are depicted as green and red lines, 
respectively. The solid, dashed, and dotted lines represent the upper bounds on 
the postrecombination gravitational waves emitted at $z_{\rm s}=1$, $10$, and $z_{\rm LSS}=1100$. 
We specifically set the experimental parameters to $\sigma_{\rm P}=0.1$\,$\mu$K-arcmin 
and $\theta=1$ arcmin. 
}
\label{OmegaGW}
\ec
\end{figure*}

In Fig.~\ref{OmegaGW}, setting the experimental parameters to $\sigma_{\rm P}=0.1$\,$\mu$K-arcmin 
and $\theta=1$ arcmin specifically, we estimate the expected upper bounds on $\Delta\Omega\rom{GW}$ 
for the postrecombination gravitational waves emitted at $z_s=1$ (solid), $10$ (dashed), and 
$z_{\rm LSS}$ (dotted), and the results are plotted as a function of the frequency
\footnote{
The upper bound $\Delta\Omega\rom{GW}$ at each $k\rom{c}$ is evaluated as an average of 
$\Delta\Omega\rom{GW}(k)$ over $k\rom{c}(1-\epsilon)\leq k\leq k\rom{c}(1+\epsilon)$. 
}. 
The left panel shows the results in the absence of the primordial gravitational waves. 
The upper bounds obtained from the curl mode become gradually tight as decreasing the frequency. 
At $k_c\sim10^{-5}$\,Mpc$^{-1}$, the constraint eventually becomes comparable to that from 
the B-mode polarization. On the other hand, the right panel shows the case including the nonzero 
primordial gravitational waves of $r=0.1$. Note that with the sensitivity of 
$\sigma_{\rm P}=0.1$\,$\mu$K-arcmin, the detection of the primordial gravitational waves via 
the lensing curl-mode is statistically less significant, i.e., $(S/N)_{\curl\curl}<2$ 
(see Fig.~\ref{snr_curl}). 
Nevertheless, the upper bounds from the curl mode almost remain the same as in the case of $r=0$, 
while those from the B-mode polarization substantially become worse. 
As a result, the curl mode gives a much tighter constraint on the long-wavelength 
gravitational waves of $k_c\lsim 10^{-3}\,$Mpc$^{-1}$. Compared to the constraint from 
the B-mode polarization, this could give a significant improvement by more than $3$ orders 
of magnitude. 
The main reason for these results comes from the fact that the contributions 
of the primordial gravitational waves give rise to a large cosmic variance leading to 
a significant impact on the large-scale B-mode polarization, while the noise contributions 
to the curl mode are basically given by the reconstruction noise which hardly changes even 
in the presence of the primordial gravitational waves. 
Another notable point in Fig.~\ref{OmegaGW} is that irrespective 
of the amplitude of the primordial contribution, the constraints on $\Delta\Omega_{\rm GW}$ from 
the curl mode have a weaker dependence on the source redshift $z_{\rm s}$ than those from the B-mode 
polarization. Hence, combining the B-mode polarization with the curl mode, we could obtain 
the stringent constraints on the postrecombination gravitational waves over wide frequency ranges 
and source redshifts.

\section{Summary} \label{sec.5} 

In this paper, based on the iterative technique for the lensing reconstruction and delensing, 
we have studied the future detectability of the gravitational-wave induced lensing from 
a high-sensitivity CMB experiment. The Fisher matrix analysis has revealed that the lensing curl-mode 
induced by the primordial gravitational waves with $r\lesssim0.1$ would be detected at more 
than $3\sigma$ significance where the consistency relation is still hard to confirm. 
As an implication of searching for the gravitational-wave induced lensing, we have considered 
the possibility to tightly constrain the postrecombination gravitational waves. We then have found that 
the lensing curl-mode is particularly sensitive to long-wavelength gravitational waves 
($k\lsim 10^{-3}\,$Mpc$^{-1}$) produced at $z\rom{s}<10$. With a high-sensitivity experiment which 
will be able to confirm the consistency relation at a few-$\sigma$ level, the lensing curl-mode 
gives a tighter constraint on their spectral energy density $\Omega\rom{GW}$,  
compared to the one from B-mode polarization by more than $3$ orders of magnitude. 

In this paper, we have assumed that the lensing B-mode polarization is generated only from 
the gradient mode and has no contributions from the curl mode. Ref.~\cite{Li:2006si} showed that 
the contributions from the curl mode to the lensing B-mode is about $<1\%$ of those from the gradient 
mode to the lensing B-mode if $r\sim 0.3$. Thus, we would safely ignore the curl-mode contributions to 
the lensing B-mode as long as we consider the primordial gravitational waves of $r\lsim 0.3$ and 
the experiment with $\sigma\rom{P}\gsim 0.05\mu$K-arcmin. In other words, such contributions have 
to be seriously taken into account for a future ultra-high-sensitivity experiment. Another concern 
in the ultra-high-sensitivity experiment would be the rotation of the polarization basis, which can 
be induced by the primordial gravitational waves  \cite{Dai:2013nda}. The contributions from 
the higher-order gradient-mode power spectrum $\mC{O}[(C_{\ell}^{\grad\grad})^2]$ may be also important 
in estimating the lensing B-mode in high-sensitivity experiments. 
These issues are left for our future work.


\begin{acknowledgments}
T.N. would like to thank Duncan Hanson and Chao-Lin Kuo for useful comments. 
This work is supported in part by a Grant-in-Aid for Scientific Research from 
the JSPS (Grant No.~26-142 for T.N., Grant No.~259800 for D.Y., and Grant No.~24540257 for A.T.). 
\end{acknowledgments}

\appendix

\bibliographystyle{mybst}
\bibliography{cite}

\providecommand{\href}[2]{#2}\begingroup\raggedright\begin{thebibliography}{10}

\bibitem{Ade:2014xna}
{\bf BICEP2}  {\bf Collaboration} , {\it ``Detection of B-Mode Polarization at
  Degree Angular Scales by BICEP2''},  {\em Phys.Rev.Lett.} {\bf 112} (2014)
  241101, [\href{http://arxiv.org/abs/1403.3985}{{\tt arXiv:1403.3985}}].

\bibitem{Adam:2014bub}
{\bf Planck}  {\bf Collaboration} , {\it ``Planck intermediate results. XXX.
  The angular power spectrum of polarized dust emission at intermediate and
  high Galactic latitudes''},  \href{http://arxiv.org/abs/1409.5738}{{\tt
  arXiv:1409.5738}}.

\bibitem{Abazajian:2013vfg}
K.~Abazajian {\em et~al.}, {\it ``Inflation Physics from the Cosmic Microwave
  Background and Large Scale Structure''},  {\em Astropart. Phys.} {\bf 63}
  (2015) 55--65, [\href{http://arxiv.org/abs/1309.5381}{{\tt
  arXiv:1309.5381}}].

\bibitem{Abazajian:2013oma}
K.~Abazajian {\em et~al.}, {\it ``Neutrino Physics from the Cosmic Microwave
  Background and Large Scale Structure''},  {\em Astropart. Phys.} {\bf 63}
  (2015) 66--80, [\href{http://arxiv.org/abs/1309.5383}{{\tt
  arXiv:1309.5383}}].

\bibitem{Hanson:2013hsb}
{\bf SPTpol}  {\bf Collaboration} , D.~Hanson {\em et~al.}, {\it ``Detection of
  B-mode Polarization in the Cosmic Microwave Background with Data from the
  South Pole Telescope''},  {\em Phys.Rev.Lett.} {\bf 111} (2013), no.~14
  141301, [\href{http://arxiv.org/abs/1307.5830}{{\tt arXiv:1307.5830}}].

\bibitem{Ade:2013gez}
{\bf POLARBEAR}  {\bf Collaboration} , {\it ``Measurement of the Cosmic
  Microwave Background Polarization Lensing Power Spectrum with the POLARBEAR
  experiment''},  {\em Phys.Rev.Lett.} {\bf 113} (2014) 021301,
  [\href{http://arxiv.org/abs/1312.6646}{{\tt arXiv:1312.6646}}].

\bibitem{Boyle:2014kba}
L.~Boyle {\em et~al.}, {\it ``On testing and extending the inflationary
  consistency relation for tensor modes''},
  \href{http://arxiv.org/abs/1408.3129}{{\tt arXiv:1408.3129}}.

\bibitem{Simard:2014aqa}
G.~Simard, D.~Hanson, and G.~Holder, {\it ``Prospects for Delensing the Cosmic
  Microwave Background for Studying Inflation''},
  \href{http://arxiv.org/abs/1410.0691}{{\tt arXiv:1410.0691}}.

\bibitem{Hirata:2003ka}
C.~M. Hirata and U.~Seljak, {\it ``Reconstruction of lensing from the cosmic
  microwave background polarization''},  {\em Phys. Rev. D} {\bf 68} (2003)
  083002, [\href{http://arxiv.org/abs/astro-ph/0306354}{{\tt
  astro-ph/0306354}}].

\bibitem{Cooray:2005hm}
A.~Cooray, M.~Kamionkowski, and R.~R. Caldwell, {\it ``Cosmic shear of the
  microwave background: The curl diagnostic''},  {\em Phys. Rev.} {\bf D71}
  (2005) 123527, [\href{http://arxiv.org/abs/astro-ph/0503002}{{\tt
  astro-ph/0503002}}].

\bibitem{Namikawa:2011cs}
T.~Namikawa, D.~Yamauchi, and A.~Taruya, {\it ``Full-sky lensing reconstruction
  of gradient and curl modes from CMB maps''},  {\em JCAP} {\bf 1201} (2012)
  007, [\href{http://arxiv.org/abs/1110.1718}{{\tt arXiv:1110.1718}}].

\bibitem{Dodelson:2003bv}
S.~Dodelson, E.~Rozo, and A.~Stebbins, {\it ``Primordial gravity waves and weak
  lensing''},  {\em Phys. Rev. Lett.} {\bf 91} (2003) 021301,
  [\href{http://arxiv.org/abs/astro-ph/0301177}{{\tt astro-ph/0301177}}].

\bibitem{Li:2006si}
C.~Li and A.~Cooray, {\it ``Weak Lensing of the Cosmic Microwave Background by
  Foreground Gravitational Waves''},  {\em Phys.Rev.} {\bf D74} (2006) 023521,
  [\href{http://arxiv.org/abs/astro-ph/0604179}{{\tt astro-ph/0604179}}].

\bibitem{Baumann:2007zm}
D.~Baumann, P.~J. Steinhardt, K.~Takahashi, and K.~Ichiki, {\it ``Gravitational
  Wave Spectrum Induced by Primordial Scalar Perturbations''},  {\em Phys.
  Rev.} {\bf D76} (2005) 084019,
  [\href{http://arxiv.org/abs/hep-th/0703290}{{\tt hep-th/0703290}}].

\bibitem{Yamauchi:2013fra}
D.~Yamauchi, T.~Namikawa, and A.~Taruya, {\it ``Full-sky formulae for weak
  lensing power spectra from total angular momentum method''},  {\em JCAP} {\bf
  1308} (2013) 051, [\href{http://arxiv.org/abs/1305.3348}{{\tt
  arXiv:1305.3348}}].

\bibitem{Saltas:2014dha}
I.~D. Saltas, I.~Sawicki, L.~Amendola, and M.~Kunz, {\it ``Anisotropic stress
  as signature of non-standard propagation of gravitational waves''},
  \href{http://arxiv.org/abs/1406.7139}{{\tt arXiv:1406.7139}}.

\bibitem{Fenu:2009JCAP}
E.~Fenu, D.~G. Figueroa, R.~Durrer, and J.~Garcia-Bellido, {\it ``Gravitational
  waves from self-ordering scalar fields''},  {\em JCAP} {\bf 10} (oct, 2009)
  5, [\href{http://arxiv.org/abs/0908.0425}{{\tt arXiv:0908.0425}}].

\bibitem{Figueroa:2013PRL}
D.~G. Figueroa, M.~Hindmarsh, and J.~Urrestilla, {\it ``Exact Scale-Invariant
  Background of Gravitational Waves from Cosmic Defects''},  {\em Physical
  Review Letters} {\bf 110} (mar, 2013) 101302,
  [\href{http://arxiv.org/abs/1212.5458}{{\tt arXiv:1212.5458}}].

\bibitem{2012:Rotti}
A.~Rotti and T.~Souradeep, {\it ``New Window into Stochastic Gravitational Wave
  Background''},  {\em Phys. Rev. Lett.} (2012).

\bibitem{Ade:2013zuv}
{\bf Planck}  {\bf Collaboration} , {\it ``Planck 2013 results. XVI.
  Cosmological parameters''},  {\em Astron. Astrophys.} {\bf 571} (2014) A16,
  [\href{http://arxiv.org/abs/1303.5076}{{\tt arXiv:1303.5076}}].

\bibitem{Lewis:1999bs}
A.~Lewis, A.~Challinor, and A.~Lasenby, {\it ``Efficient Computation of CMB
  anisotropies in closed FRW models''},  {\em Astrophys. J.} {\bf 538} (2000)
  473--476, [\href{http://arxiv.org/abs/astro-ph/9911177}{{\tt
  astro-ph/9911177}}].

\bibitem{Smith:2010gu}
K.~M. Smith {\em et~al.}, {\it ``Delensing CMB Polarization with External
  Datasets''},  {\em JCAP} {\bf 1206} (2012) 014,
  [\href{http://arxiv.org/abs/1010.0048}{{\tt arXiv:1010.0048}}].

\bibitem{Wu:2014hta}
W.~Wu {\em et~al.}, {\it ``A Guide to Designing Future Ground-based Cosmic
  Microwave Background Experiments''},  {\em Astrophys.J.} {\bf 788} (2014)
  138, [\href{http://arxiv.org/abs/1402.4108}{{\tt arXiv:1402.4108}}].

\bibitem{Okamoto:2003zw}
T.~Okamoto and W.~Hu, {\it ``CMB Lensing Reconstruction on the Full Sky''},
  {\em Phys. Rev.} {\bf D67} (2003) 083002,
  [\href{http://arxiv.org/abs/astro-ph/0301031}{{\tt astro-ph/0301031}}].

\bibitem{Dai:2013nda}
L.~Dai, {\it ``Rotation of the cosmic microwave background polarization from
  weak gravitational lensing''},  {\em Phys.Rev.Lett.} {\bf 112} (2014), no.~4
  041303, [\href{http://arxiv.org/abs/1311.3662}{{\tt arXiv:1311.3662}}].

\end{thebibliography}\endgroup

\end{document}